# Science In the Cloud (SIC): A use case in MRI Connectomics

Gregory Kiar[1,2], Krzysztof J. Gorgolewski[3], Dean Kleissas[4], William Gray Roncal[4,5], Brian Litt[6,7], Brian Wandell[3,8], Russel A. Poldrack[3], Martin Wiener[9], R. Jacob Vogelstein, Randal Burns[5], Joshua T. Vogelstein[1,2]

Corresponding Author: Joshua T. Vogelstein `jovo@jhu.edu`

**Abstract**

Modern technologies are enabling scientists to collect extraordinary amounts of complex and sophisticated data across a huge range of scales like never before. With this onslaught of data, we can allow the focal point to shift from data collection to data analysis. Unfortunately, lack of standardized sharing mechanisms and practices often make reproducing or extending scientific results very difficult. With the creation of data organization structures and tools which drastically improve code portability, we now have the opportunity to design such a framework for communicating extensible scientific discoveries. Our proposed solution leverages these existing technologies and standards, and provides an accessible and extensible model for reproducible research, called "science in the cloud" (SIC). Exploiting scientific containers, cloud computing, and cloud data services, we show the capability to compute in the cloud and run a web service that enables intimate interaction with the tools and data presented. We hope this model will inspire the community to produce reproducible and, importantly, extensible results which will enable us to collectively accelerate the rate at which scientific breakthroughs are discovered, replicated, and extended.

# 1 Introduction

Neuroscience is currently in a golden age of data and computation. Through recent technological advances [1], experimentalists can now amass large amounts of high quality data across essentially all experimental paradigms and spatiotemporal scales; such data are ripe to reveal the principles of brain function and structure. In fact, many public datasets and open-access data hosting repositories are going online [2; 3].

Concurrent with this onslaught of data is a desire to run analyses, not just on data collected in a single lab, but also on other publicly available datasets. Various tools have been developed by the community which solve a wide variety of computational challenges on all types of data, enabling difficult scientific questions to be answered. With the ability to perform analyses often dependent only upon access to data and code resources, neuroscience is now more accessible, with a lower barrier to entry.

However, there is no tool or framework that enables research to be performed and communicated in a way that lends itself to easy extensibility, much less reproducibility. Currently, re-performing and extending published analyses whether through data or code is often unbearably difficult: (i) data may be closed-access; (ii) data may be organized in an ad hoc fashion; (iii) the code may be closed-source or undocumented; (iv) code may have been run with undocumented parameters and dependencies; (v) analyses may have been run with code compiled for specific hardware. These properties make validating and extending scientific claims challenging.

A focus on reproducibility is already commonplace in a variety of disciplines. In genomics, Bioboxes [4] provide a framework for reproducible and interchangeable analysis containers, and tools are exploiting scalable computing solutions and being published with reproduction instructions (see: [5; 6]). Commentaries on reproducible research provide suggestions to researchers on how to tackle the challenges that are present in their scientific settings [7; 8]. While these works have accelerated reproducibility and extensibility in their fields, the methods proposed do not scale to the cloud or enable real-time interactivity, and have yet to be thoroughly applied to the burgeoning field of computational neuroscience.

The notion of a universally web-viewable laboratory [9] is also growing in popularity, and many initiatives have been successful in contributing to this vision. In plant biology, CyVerse [10] provides infrastructure for tools, data, and education. In neuroscience, platforms such as LONI's Pipeline [11] and neuGRID [12] alleviate the burden of managing captive computing resources and integrating them with datastores, while NeuroDebian [13] provides quick and easy access to a variety of neuroimaging tools. Leveraging the NeuroDebian platform, NITRC has encouraged a transition to the cloud by releasing an AMI[1] preloaded with commonly used packages. In parallel, many groups have strived to breach the frontier through such efforts as developing sophisticated resource estimation-based deployment strategies [14], and these have shown the great potential for a cloud-based approach to neuroimaging [15]. Each of these projects has made valuable contributions to the progress towards accessibility and portability of neuroscience research.

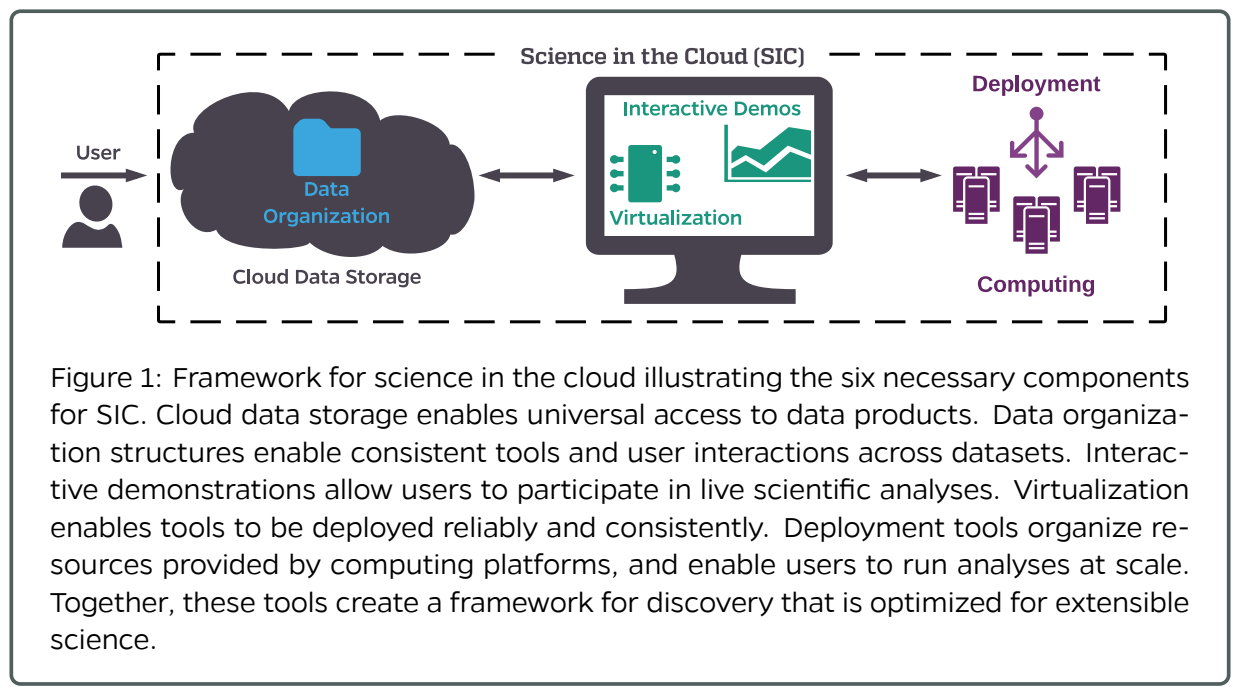

Figure 1: Framework for science in the cloud illustrating the six necessary components for SIC. Cloud data storage enables universal access to data products. Data organization structures enable consistent tools and user interactions across datasets. Interactive demonstrations allow users to participate in live scientific analyses. Virtualization enables tools to be deployed reliably and consistently. Deployment tools organize resources provided by computing platforms, and enable users to run analyses at scale. Together, these tools create a framework for discovery that is optimized for extensible science.

We propose a solution to these gaps in the form of a framework which leverages publicly documented and deployable cloud instances with specific pipelines installed and configured to extend published findings: an implementation we simply term "science in the cloud," or, SIC (Latin for "thus was it written"). SIC instances have several fundamental components, as summarized in Figure 1. To address data access, we put data in the cloud. To address data organization, we utilize recently proposed data standards. To address closed source and undocumented code, we generate open-source code and interactive demonstrations. To address software and hardware dependencies, we utilize virtualization, automated deployment, and cloud computing. SIC puts these pieces together to create a computing instance launched in the cloud, designed not only for generating reproducible research, but also enabling easily accessible and extensible science for everyone. SIC is designed to minimize the bottlenecks between publication and novel discoveries; leveraging the experience of the community, we propose a solution for transitioning to a universal, and "future-proof," deployment of software to the cloud.

We introduce and document an example use case of SIC with the ndmg pipeline, thus entitled SIC:ndmg. We have developed a capability which enables users to launch a cloud instance and run a container which performs an analysis of a cohort of structural and diffusion magnetic resonance imaging scans by (i) downloading the required data from a public repository in the cloud, (ii) fully processing each subject's data to estimate a connectome for each subject's associated graph statistics, and, optionally, (iii) plot quality control figures of various multivariate graph statistics.

## 2 Methods

There are six key decisions which must be made when following SIC: data storage, data organization, interactive demonstrations, virtualization, deployment, and computing. The selection made for each of these components will have a significant impact on available selections for the others. The final product will be a highly interdependent network of tools and data. Table 1 shows a summary of the selections made for each of the criteria enumerated in the previous section with rationales for the decisions. In general, the tools selected were those which provided the most command-line/Application Programming Interface (API) support for their service and had the most complete documentation or online support community, enabling setup with relative ease.

**Cloud Data Storage**  There are several options when storing data in a publicly accessible location, such as a cloud storage service or public repositories. Depending on the nature of the data being stored, different concerns (such as privacy) must be satisfied. For instance, sensitive data (i.e. not anonymized/de-identified) requires authentication for access, whereas de-identified data does not. It is our recommendation to host de-identified data in the cloud and store linking metadata privately on HIPPA (or equivalent) compliant organization datastores. Researchers who may not wish to release their data prior to publication are encouraged to store their data with secure protocols. The datastore should also be accessible through an API, or another interface enabling developers to access the data programmatically. Depending on the desired organization, autonomy is also a valuable feature, affording the developer full control on how the data is stored, as opposed to working within the confines of an existing infrastructure. The type of virtualization (described below) used may also influence the types of shared datastores which will be natively compatible with the application. Considering the above, Amazon's S3 service was used in this SIC implementation because it satisfied all of these requirements. While Google's Cloud Engine or Microsoft Azure also satisfy these requirements, the decision to use S3 was made based upon our existing domain knowledge and familiarity with each of these systems.

**Data Organization** The newly publicly-available data then needs to be organized in accordance with a data specification which enables users to navigate the repository successfully. Such standards include both file formats, which can be interpreted by programs, as well as folder organizations, which enable grouping of data by subject, observation, type, etc. Depending on the modality of data being used, there are different structures which can be adopted. In the case of MRI, the BIDS [16] specification is a well-documented and community-developed standard which is intuitive and allows data to be both easily readable by humans and navigated by programs. Organizations such as "Neurodata without Borders" [17] would serve as additional options for physiology data, but are unsuitable for this application. Formats such as MINC [18] focus heavily on metadata management but less on file hierarchy, making them useful though not fully sufficient for this application. Though some standards may consider securely handling identifying information, we recommend only storing de-identified data publicly to avoid possible security risks.

Table 1: There are six key components which must be selected for SIC. **Bold** indicates the selections made here, with their positive and negative qualities compared to some alternatives.

| Hurdles | Available Tools | Pros of Selection | Cons of Selection |
| --- | --- | --- | --- |
| 1) Data Storage | **S3**, Dropbox, Google Drive | API, pay-by-usage | requires familiarity with Amazon tools |
| 2) Data Organization | **BIDS** [16], NWB [17], MINC [18] | documented, validator, active community | new, not yet fully adopted |
| 3) Interactive demo's | **Jupyter**, R Notebook, Shiny | versatile, accessible | optimized for Python |
| 4) Virtualization | **Docker**, Virtualbox [19], VMware [20] | lightweight, self-documented | -- |
| 5) Deployment | **Batch/ECS**, Kubernetes [21], MyBinder [22], CBRAIN [23], Nextflow [24] | no additional dependencies | restricted to Amazon's cloud |
| 6) Computing | **EC2**, Google Compute Engine [25], Microsoft Azure [26] | scalable, flexible | requires technological expertise |

**Interactive Demonstrations**   To encourage use of data and the tools used to analyze it, interactive demonstrations that enable users to visualize and work with some subset of the data are extremely valuable. Various programming languages have different types of demonstration environments available which either enable full interactivity or are pre-compiled to display code and results. A popular tool for interactive development and deployment of Python code is Jupyter, and thus was the tool used here. The popularity of this tool hopefully increases the average user's familiarity with the interface, lowering the barrier to entry for interacting with SIC:ndmg. If a developer is more familiar with another programming language, there is no particular reason why one would select Jupyter over an equivalent package in R, such as R Notebook.

**Virtualization**   Developing and distributing virtualized environments containing all necessary code products guarantees consistent dependencies and application setup, and therefore minimizes user effort to obtain expected performance. These virtual environments should be able to be deployed on any operating system and have minimal hardware-dependent code. A key desiderata is that the virtualization system minimizes unnecessary overhead for the application. Though it does not affect run-time performance, a repository of public machine images is an attractive feature for this model as it enables sharing configurations. Docker [27] was chosen because it satisfies these practical requirements, and the accessibility of Docker Hub enables images to be quickly found and deployed. Virtual machines such as those created in Virtual Box [19] or VMware [20] provide lots of range in terms of operating systems which can be launched and allow native access to the machine through a GUI. However, though these are great features, they are unnecessary for this application. An additional attractive feature of Docker is that translating a `README` file (which enumerates dependencies or installation instructions) to a Dockerfile forces developers to improve their documentation and increases the useability of their tool. Though this is certainly extra work for the developer, the process requires only knowledge of the documented Docker schema and the editing of plain-text files, which we believe to be a relatively low cost to the developer.

**Deployment**   Deployment platforms allow users to define a specific set of instructions that can be launched on a single machine or multiple machines simultaneously. In physical hardware configurations, a cluster's scheduler would play this role; in the cloud, such tools are able to take advantage of computing resources across different locations and services, and enable scaling with the amount of processing required. Middleware such as Kubernetes [21], Tutum[2], or Nextflow [24] can enable a user to distribute their jobs across a cluster existing in different computing environments (i.e. separate clouds). When using a single cloud, such as Amazon or Google, native applications support managing resources efficiently. In the case of SIC:ndmg, we elected to deploy entirely in Amazon's cloud; therefore, we used Amazon's Batch to launch the pipeline distributed across multiple computing nodes, and Amazon's ECS to deploy a distributed and scalable SIC service. Tools such as CBRAIN [23], LONI [11], and

**NeuroData**

MyBinder [22] also enable distributed deployment of code, but are more specialized in the requirements of the tools and services that can be launched and are thus more restrictive.

**Computing** Cloud computing services enable users to launch customized machines with specific hardware configurations and specifications, making them versatile for different varieties and scales of analyses. The more general the hardware that can be used, the more accessible the tool is for a user to adapt and use in their own environment. Selecting the commercial cloud for deployment as opposed to data center resources enables greater accessibility and transparency to users, is more scalable, and enables parallel jobs to be run in completely isolated resources. Cloud deployments also provide consistent performance across nodes, and have a much lower start-up cost than utilizing local computing resources. Since there were no specific hardware requirements in this application, and there existed previous in-house experience with the service, Amazon's EC2 was selected in this usecase. The benefit of using EC2 is that deploying code at different scales and locations is trivially extendable, so implementations can be easily taken from prototype to deployment. Amazon's cloud enables launching computing resources based on Amazon Machine Images (AMIs) with preinstalled dependencies, increasing the flexibility of the processes which can be launched.

Further details of our specific implementation and methods are provided in Appendix A.

# 3 Results

We demonstrate a working example of SIC, SIC:ndmg. The ndmg pipeline [28] is an open-source, scalable pipeline for human structural connectome estimation from diffusion and structural MR images (collectively referred to hereafter as "multimodal MRI", or M3RI for brevity). The result is a portable and easily extensible tool for scalable connectome generation. A live demonstration is presented that enables reader interaction with the pipeline at the cost of a simple URL click, and data products of the tool are presented in both the context of `reproducibility' and `extensibility.' This tool enables quantitative structural analyses of the human brain to be performed on populations of M3RI scans, and can lead to discoveries of the relationship between brain connectivity and neurological disease.

## 3.1 Neuroscience as a Service

The analysis transforms "raw" M3RI data into graphs. Kiar et al., (in preparation) describes the pipeline in detail; here we provide a brief overview. The pipeline (Figure 2) consists of four main steps: registration, tensor calculation, tractography, and graph generation. Note that the choices below are made for expediency and simplicity; other choices might be beneficial depending on context. Table 2 summarizes the duration and cost of each step for a given dataset processed and stored in the cloud.

Registration in ndmg is performed in several stages using FSL [29]. First, the diffusion image is self-aligned and noise-corrected using the `eddy_correct` function. Second, the transform is computed which aligns the B0 volume of the diffusion image to the structural scan using `epi_reg`. Third, the transform between the structural image and a reference atlas is computed with `flirt`. Finally, the transforms are combined and applied to the self-aligned

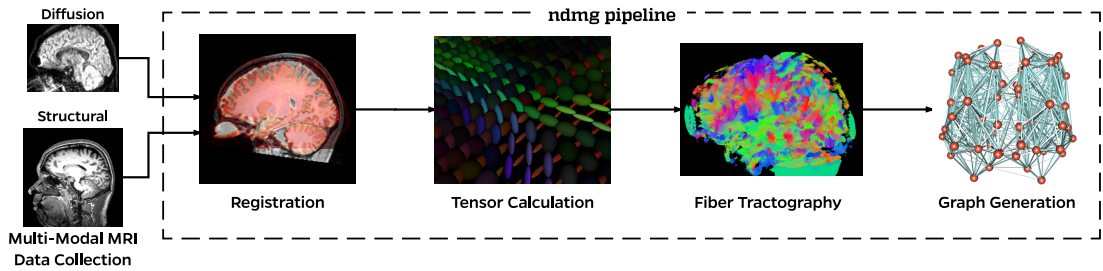

Figure 2: Structure of the ndmg pipeline connectome estimation. Taking as input diffusion and T1 weighted MRI, ndmg first aligns the diffusion data to a reference atlas by means of the T1 image. Tensors are then computed from the aligned diffusion volume. Fiber streamlines are generated by performing tractography on the tensors. Finally, the fibers are mapped between regions of interest (ROIs) which then become nodes in the graph.

diffusion image. The tensor calculation and tractography steps are performed with the DiPy package [30]. A simple tensor model fits a 6-component tensor to the image, and deterministic tractography with the EuDx algorithm is run, producing a set of streamlines. Graph generation takes as input the fiber streamlines, and maps them to regions of interest (ROIs) defined by a pre-built parcellation (such as those packaged with FSL or generated with brain segmentation algorithms) and returns an ROI-wise connectome. An edge is added to the graph for each pair of nodes along a given fiber. The final step is computing (multivariate) graph statistics on the estimated connectomes. The statistics computed are [31]: number

Table 2: Approximate cost and time breakdown per subject of the ndmg pipeline running in Amazon EC2 with data stored in S3 and computation with m4.large machines at spot pricing of $0.0135 per hour (Accessed on 2017/01/04). The values were obtained by processing data from the NKI1 dataset with 40 sessions. The reader should note that Amazon S3 data I/O is not free, as it may appear, but is simply inexpensive for data this size.

| Operation | Time per session (min) | Cost per session (1/100 USD) |
| --- | --- | --- |
| data storage | -- | 1.048/month |
| data I/O | -- | 0.000 |
| **Total** | -- | **1.048/month** |
| registration | 25 | 0.563 |
| tensor calculation | 2 | 0.045 |
| fiber tractography | 5 | 0.112 |
| graph generation | 30 | 0.675 |
| **Total** | **62** | **1.395** |

NeuroData

of non-zero edges, degree distribution, eigen sequence, locality-statistic 1, edge weight distribution, clustering coefficient, and betweenness centrality. These statistics provide insight into the structure of the brain graphs, and provide a low-dimensional feature by which the graphs for different scans can be compared to one another. To provide a preliminary quality control step, we plot the graph statistics [31] for each graph (Figure 4).

## 3.2 Live Demonstration

A demonstration of SIC:ndmg is available at http://scienceinthe.cloud/. This SIC instance is deployed via ECS on an Amazon micro-instance which is very affordable, so it can stay online indefinitely with little cost or maintenance ($100/year). This instance is running a Jupyter server which contains the demonstration notebook, sic_ndmg.ipynb. Launching the notebook pulls up an interface which resembles that of Figure 3A.

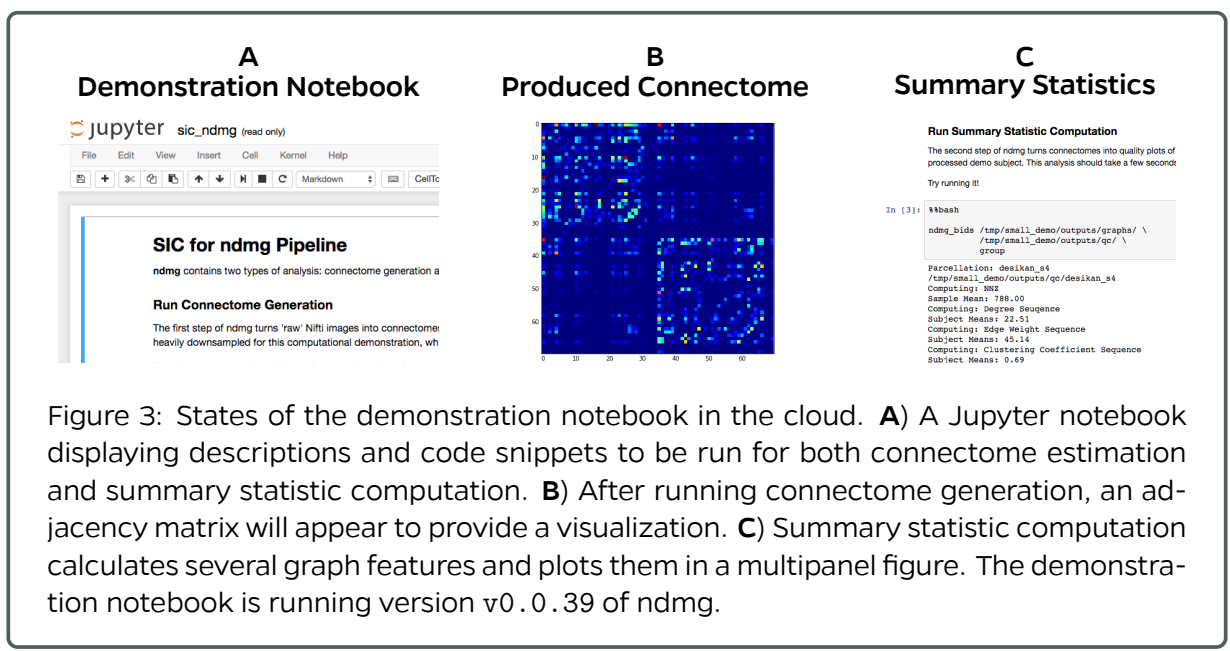

Figure 3: States of the demonstration notebook in the cloud. **A**) A Jupyter notebook displaying descriptions and code snippets to be run for both connectome estimation and summary statistic computation. **B**) After running connectome generation, an adjacency matrix will appear to provide a visualization. **C**) Summary statistic computation calculates several graph features and plots them in a multipanel figure. The demonstration notebook is running version v0.0.39 of ndmg.

For demonstration purposes, a downsampled subject is used in this notebook which reduces analysis time from ∼1 hr/subject/core to ∼3 min/subject/core. The ndmg pipeline has two levels of analysis: graph generation and summary statistic computation. Graph generation is the process of turning diffusion and structural MR images into a connectome (i.e. brain graph), and the summary statistic computation produces a graph of several graph features on each produced connectome and plots them together. Running through the notebook (Figure 3A) chronologically will produce the brain graph, display the graph (Figure 3B), compute summary statistics (Figure 3C), and then plot the statistics.

## 3.3 Reproducible Results

In addition to the live demonstration, SIC:ndmg was used to process the NKI1 [32] dataset consisting of 40 M3R scans. Instructions on setting up a cluster and running this analysis can be found in Appendix A. The NKI1 dataset is made publicly available through CORR [32], but has been organized in accordance to the BIDS [16] specification and re-hosted on our public S3 bucket, `mrneurodata`. The dataset consists of MPRAGE, DWI, and fMRI scans, where each subject has been scanned at least twice for each modality. More information about the subjects in this dataset and the scanning parameters used can be found on the CORR website[3].

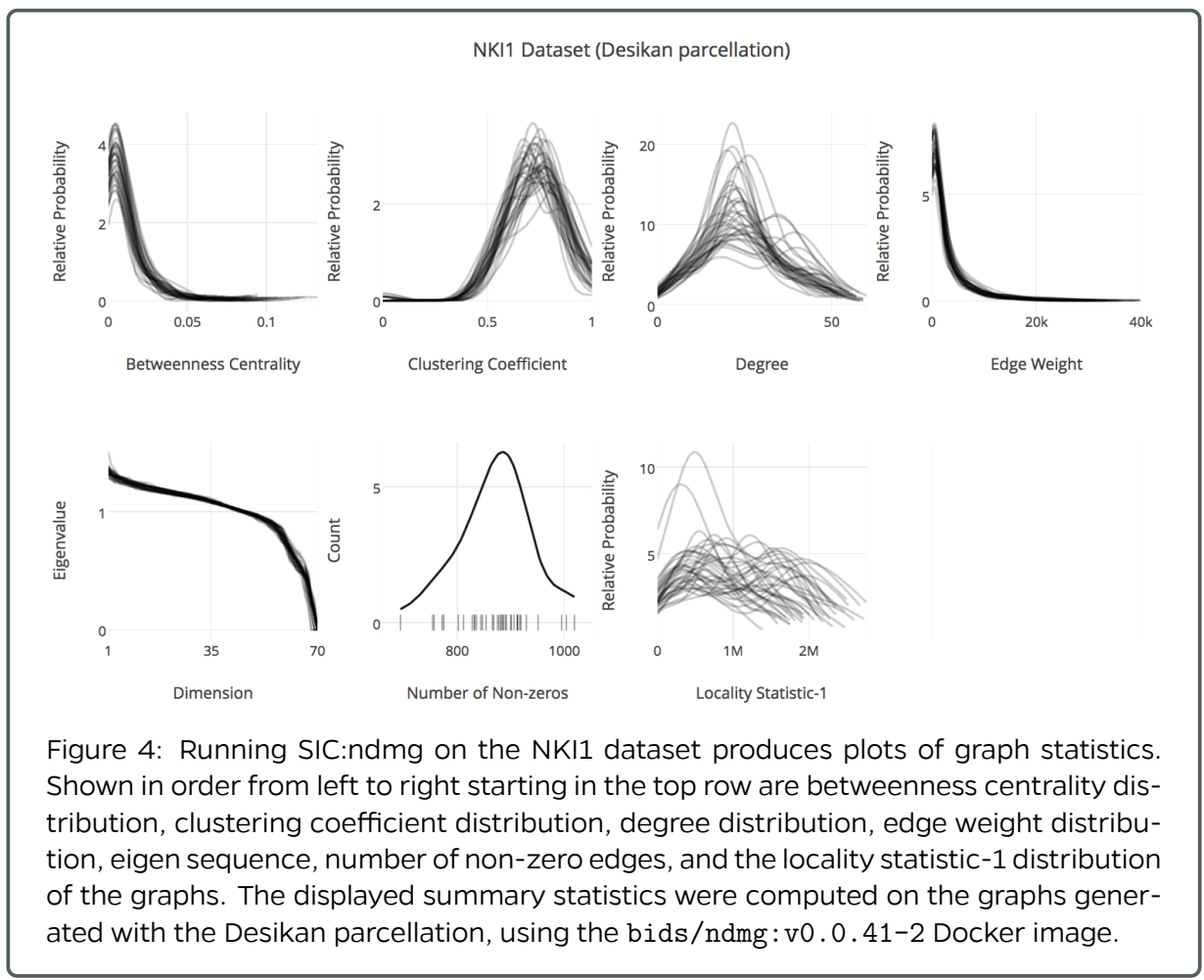

Figure 4: Running SIC:ndmg on the NKI1 dataset produces plots of graph statistics. Shown in order from left to right starting in the top row are betweenness centrality distribution, clustering coefficient distribution, degree distribution, edge weight distribution, eigen sequence, number of non-zero edges, and the locality statistic-1 distribution of the graphs. The displayed summary statistics were computed on the graphs generated with the Desikan parcellation, using the `bids/ndmg:v0.0.41-2` Docker image.

Running the Docker-hosted scientific container `bids/ndmg:v0.0.41-2` on the NKI1 dataset produced Figure 4, costing under $1, as is summarized in Table 2. Table 3 summarizes the parameters used as inputs to SIC:ndmg to generate the graphs. Figure 4 provides insight into the variance of the dataset through a variety of different metrics. According to published work on these summary statistics [31], this dataset and pipeline combination produces expected results. A key benefit of this visualization is that it has high information density, showing us distributions for a variety of features for a large number of graphs, as opposed to more common 1-dimensional features [33]. This figure was produced by the parameters summarized in Table 4.

The demonstration in the previous section executed the exact same pipeline that was used to generate Figure 4. The sole difference between execution of the demonstration and this implementation -- aside from the data being processed -- is the specific Docker container being used. The reason for this difference is that the demonstration is required to run as a web service, so additional packages and setup are required.

Table 3: Command line arguments for connectome generation

| Parameter | Value |
| --- | --- |
| data input directory | `/data/raw` |
| data output directory | `/data/connectome` |
| analysis level | `participant` |
| bucket name | `mrneurodata` |
| path on bucket | `NKI24` |

Table 4: Command line arguments for summary statistic computation.

| Parameter | Value |
| --- | --- |
| data input directory | `/data/connectome/graphs` |
| data output directory | `/data/qc` |
| analysis level | `group` |

### 3.4 Extensible Results

A crucial property of SIC is the simplicity it affords users to perform extensible science. Extensibility in this context can occur on several levels, including changing or adding (i) data, (ii) analyses, or (ii) visualizations. Figure 5 shows an example of such extensibility. A different dataset, the KKI2009 dataset [34], was processed using modified code, plotting the degree distribution on a log scale, with an additional plot added for cumulative variance analysis. The container used for this analysis on Docker hub is `bids/ndmg:v0.0.41-2`. Further details and instructions about how to extend SIC:ndmg specifically are available in Appendix B.

## 4 Discussion

Though the exemplar application used to demonstrate the value of SIC was the one-click ndmg pipeline, the framework is not restricted to this tool, or even one-click tools at all. For instance, a recent manuscript presented the notion of BIDS Apps [35]: containerized neuroimaging applications which operate on data stored in the BIDS data structure. These apps[4] enable complex workflows to be executed, often taking in configuration files to allow for complicated parameter sets to be delivered more conveniently than via the command line. Such containers are a terrific usecase for SIC, and can be seamlessly interchanged with one another in a given deployment. SIC can use tools such as FreeSurfer or ANTs in certain process-ing steps with no software changes. Developing pipelines within the SIC framework enhances their reproducibility and the extensibility of publications using them, potentially increasing their scientific impact.

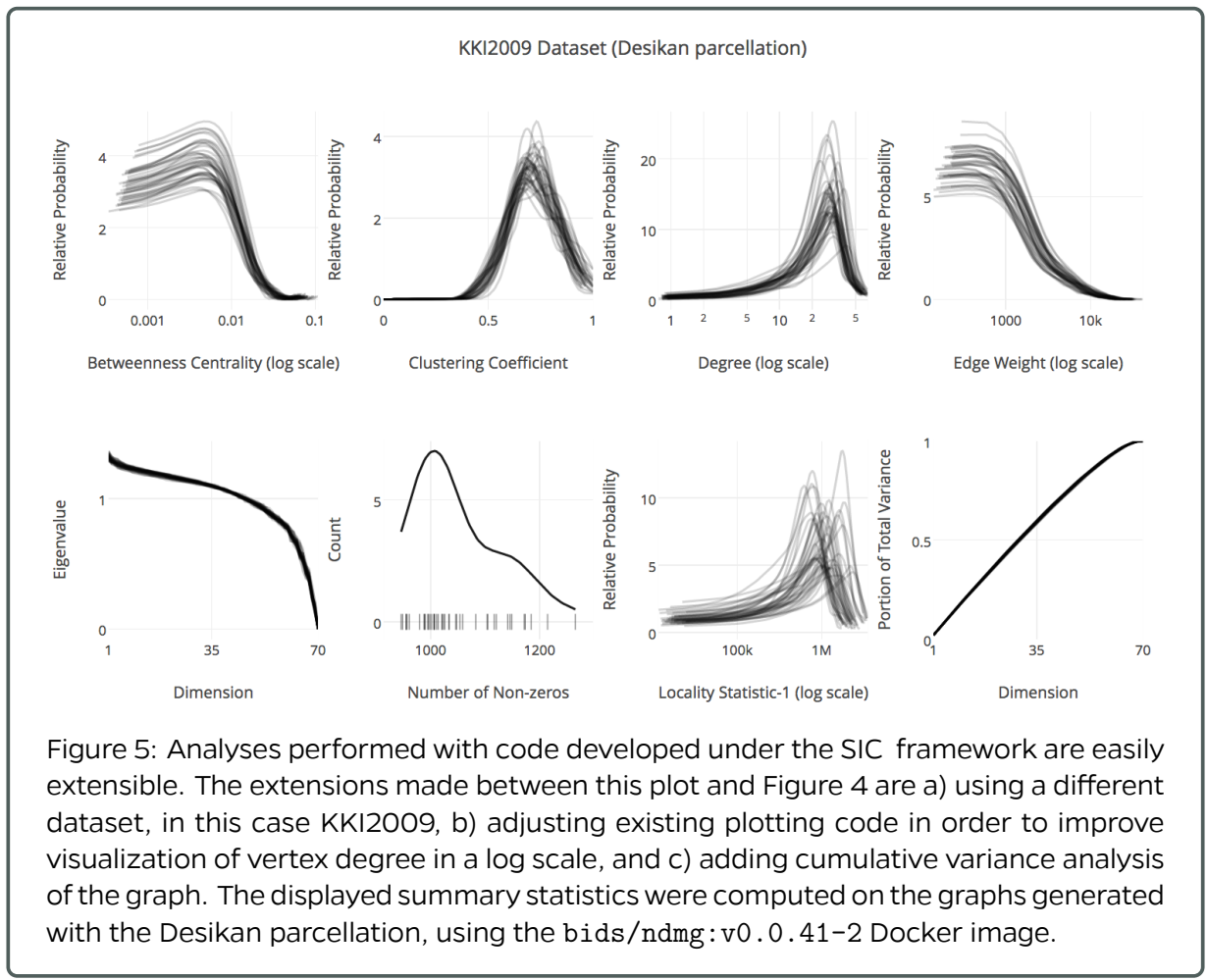

Figure 5: Analyses performed with code developed under the SIC framework are easily extensible. The extensions made between this plot and Figure 4 are a) using a different dataset, in this case KKI2009, b) adjusting existing plotting code in order to improve visualization of vertex degree in a log scale, and c) adding cumulative variance analysis of the graph. The displayed summary statistics were computed on the graphs generated with the Desikan parcellation, using the `bids/ndmg:v0.0.41-2` Docker image.

The SIC framework does not need to be confined to monolithic tools and containers. With further work, this concept can be integrated into a platform in which users are able to launch a variety of analyses on a variety of datasets. The self-documenting and reproducible web-calls which launch cloud containers performing computational tasks have potential to drastically improve the feedback loop between a scientist and their peers. This enables analyses to be easily replicated and refined, thus expediting scientific discovery. Tools such as Binder [22] accomplish this beautifully for Python, but the benefits of SIC are that this model can be applied not only to any containerizable application, but big data as well.

The distinct advantage of using Docker for virtualization as opposed to virtual machines is the lack of both computational and data overhead. Though virtual machines can be used for pipeline deployment, they are based upon hard drive files which can bloat the host system. Virtual machines also require computational overhead to distribute processes to the host system, which Docker interfaces with directly. In many applications, virtual machines are a wise or even necessary tool of choice, though when the sole objective is the execution of a pipeline followed by termination of the environment, the benefits of minimal overhead often outweigh those of the additional features which may be available through virtual machines. Tools which aid in the deployment of virtualized environments such as Vagrant can be paired with a method of virtualization, whether Docker or otherwise, and they provide further documentation describing the process for launching an environment containing a given tool for execution.

The selections made in SIC:ndmg regarding the six technological components highlighted above were chosen based on what the authors perceived to be most widely used and supported in the active online community. Other tools enumerated in Table 1 provide alternative features which can make SIC instances appear and run quite differently when developed separately, but ultimately provide a comparable experience for the user. For instance, the decision to store data independently from a public repository (such as NITRC [36], LONI's IDA [37], LORIS [38], or ndstore [39]) leaves the onus of data organization on the developer rather than the repository, but in either case the user is able to access the data they need. This decision in particular was made so that the developer would have complete control over their data and implementation. However, hosting data within an environment such as those listed would have the advantage of enabling use of the infrastructure already built to support these platforms, such as performing meta-analyses and tracking provenance of the data itself, and is an exciting avenue for future work. While functionality for deploying in parallel to the cloud was developed with Amazon's Batch directly for interfacing with their cloud, alternative deployment tools such as Kubernetes are attractive options, because they provide clear visualizations of running processes and process versions and would enable SIC to deploy pipelines across multiple computing clouds or clusters. Deployments making use of local datacenters as opposed to the cloud are identical in execution to those in the cloud, once Docker (or the virtualization engine of choice) is installed on the shared resources and a scheduling framework is available.

This manuscript proposes a model for extensible and accessible development that did not strain those who have already been developing or using reproducible tools, but rather enhanced their ability to do so. Domain knowledge, such as that of Docker, is not uniform across disciplines, and this may discourage developers from complying with this methodology. However, it is our belief that the proposed framework does not require additional development beyond what already goes into creating and using a reproducible tool. For instance, in the case of Docker, a Dockerfile simply documents the instructions which are to be executed upon booting a brand-new computer and installing a given tool and its dependencies. Documenting this process is essential for developers, and many tools contain a `README` file describing the installation process. Once a Docker container exists, the process of re-executing and testing these instructions often requires far fewer keystrokes and ambiguity in the instructions is eliminated. There are certainly start-up costs when transitioning to new tools such as virtualization platforms, but it is our view that the gained transparency and portability within SIC greatly outweighs the costs.

In summary, the SIC framework presents a standard of reliability and extensibility for scientific data distribution and analysis. SIC is an important building block towards a global scientific community, regardless of scientific discipline, and provides a practical implementation of the idiom that science is done by "standing on the shoulders of giants."

# Acknowledgements

This project stemmed from a sequence of three different initiatives. First, the Global Brain Workshop[5] brought together a collection of $60+$ scientists who converged on a set of grand challenges for global brain sciences. There was universal agreement that a global framework [40] would be instrumental in transitioning neuroscience from a data deluge to a data delight. Then, at the Open Data Ecosystem for Neurosciences[6], the working group on reproducibility decided that an example of a reproducible and extensible framework would be highly informative for ourselves and the greater community. Finally, the inaugural Stanford Center for Reproducible Neuroscience Coding Sprint[7] brought leaders in neuroimaging from around the globe to chart a path forward with standardizing a process for containerizing both open- and closed-source tools [35].

The authors would like to graciously thank: NIH, NSF, DARPA, IARPA, Johns Hopkins University, and the Kavli Foundation for their support. Specific award information can be found at https://neurodata.io/about.

# Affiliations

[1]Department of Biomedical Engineering, Johns Hopkins University, Baltimore, MD, USA.
[2]Center for Imaging Science, Johns Hopkins University, Baltimore, MD, USA.
[3]Department of Psychology, Stanford University, Stanford, CA, USA.
[4]Johns Hopkins University Applied Physics Lab, Columbia, MD, USA.
[5]Department of Computer Science, Johns Hopkins University, Baltimore, MD, USA.
[6]Department of Bioengineering, University of Pennsylvania, Philadelphia, PA, USA.
[7]Department of Neurology, Hospital of the University of Pennsylvania, Philadelphia, PA, USA.
[8]Center for Cognitive and Neurobiological Imaging, Stanford University, Stanford, CA, USA.
[9]Department of Psychology, George Mason University, Fairfax, VA, USA.

# Declarations

**Competing Interests**   The authors declare no competing interests in this manuscript.

NeuroData

# Appendix A  Reproduction Instructions

Outlined here are the required steps to reproduce both the analysis of data in the cloud, as well as the live demonstration notebook server. In the command blocks which follow, all commands preceded by a $ should be executed. Commands which are executed in a single line but were too long to fit on the page end with \ and are carried over to lines which have been indented. Below, the assumption is that the commands are being executed on a Unix-based machine with access to a terminal. If one is working with a Windows operating system, installing a GNU environment such as Cygwin[8] will enable the user to have a similar experience.

## A.1  Processing Data in the Cloud

Through use of the AWS Batch tool, a scalable computing cluster is able to be launched in the cloud and jobs can be submitted to it for analysis via the command line. The process which must be followed is: create a computing environment, create a job-submitting queue, create a job definition, and finally, submit jobs to the cluster. We discuss how to accomplish each of these steps, and provide the scripts which were used for the deployment presented in this manuscript. One prerequisite for the instructions that follow is that the data in question for processing is made available at a public read- and write-able S3 bucket in the BIDS data format.

### A.1.1  Setting up an AWS Batch cluster

Following the AWS Batch[9] Getting Started tutorial, one can create a cloud computing cluster for themselves, establish a job-accepting queue, define jobs, and submit jobs to the queue, all within the web console. Though these operations can be done via the command line as well, they will only need to be performed once so it is not significantly advantageous to script these steps.

At each of these steps there are several decisions which must be made regarding the size of the cluster, the number of cores, what container image to use in your job definition, and more. The definitions used to setup the ndmg pipeline and cluster can be found in the SIC Github repository[10].

### A.1.2  Launching jobs on the cluster

Once the cluster is live and a job definition for the ndmg pipeline has been created, jobs can start being submitted to the queue. When submitting a job to the cluster, one must first take the existing task definition for the process they are trying to run, and then override relevant portions of this definition for the desired usecase. For instance, if one wishes to run a single subject from the NKI1 dataset stored on our public S3 bucket, they may create a job submission which summarizes this[11]. This step can be done either from within the console or via the command line. In order to use the command line interface, one must first install the Amazon CLI tool and configure it with their user credentials to ensure that processes launched via the command line and web console are linked.

If one wishes to launch many jobs at once, the ndmg package contains a script which accepts an S3 bucket, a path to the dataset on that bucket, and will then launch all of the subjects within that dataset on the previously created cluster. Currently, this functionality does not exist within the Docker container version of ndmg, as it requires supplying authentication information to Amazon. However, passing this information to the Docker container safely and securely is a feature which the developers hope to eventually make available. To use this script, one must have installed the ndmg package in Python, and then may type the following line from a terminal window:

```
$ ndmg_cloud --bucket s3_bucket_name --bids_dir \
    path_on_bucket
```

As well as receiving output to the terminal, opening the Batch web console to view that the jobs have been launched can serve as confirmation that this is completed. Once the processing is complete, the outputs will be pushed back to the provided S3 bucket and the results can be analyzed.

## A.2    Launching Demonstration Notebook Service

The interactive SIC:ndmg notebook can be a valuable way to experience the ndmg pipeline and walk through the steps it takes, from generating graphs to plotting them and producing summary statistics. This interactive notebook is contained within its own Docker container, and automagically launches the service upon creating an instance of the container. We will walk through the brief process of launching this container on your local machine so that you may interact with it or change it yourself.

### A.2.1    Setting up Your Machine

The only required setup for running locally is to install Docker. Docker has installation helpers for all operating systems available on their website[12]. Once Docker is installed, it is important to make sure that the port 8888 is open for Docker. In the case of Mac OS X and Linux, this should be the case automatically, but for Windows it currently must be opened through the networking options of VirtualBox.

### A.2.2    Launching the Docker container

The user can launch the service with a single command from a terminal with access to Docker. This terminal is the standard terminal on Linux or Mac OS X, and can be the Powershell or provided terminal when installing Docker. The following command launches this service:

```
$ git clone https://github.com/neurodata/sic ~/sic
$ cd ~/sic/code/jupyter
$ docker build -t neurodata/sic .
$ docker run -d -p 8888:8888 neurodata/sic
```

You can interact with the demo via a web browser. Navigate to `localhost:8888` in the broswer of your choosing to see this service live.

17                                                                                          NeuroData

# Appendix B  Extension Instructions

As this is a living and breathing project undergoing development, changes are being made regularly. The reproduction instructions given in Appendix A will reproduce the exact results presented within this manuscript. There are several ways described below which enable staying up-to-date with the project and performing one's own analyses using this tool.

## B.1  Updating the ndmg Container

In order to achieve state-of-the-art performance from the ndmg pipeline, the version of the container being used should be updated to the `latest` release. In the job definition created above, specifying that the container image being used is `bids/ndmg:latest` as opposed to `bids/ndmg:v0.0.41-1`, for instance, will ensure that the most recent version of the code is being used.

## B.2  Using Your Data

The ndmg pipeline processes data according to the BIDS data specification. To use the tool with an alternate dataset, it first needs to be organized according to this specification. This can be validated using the BIDS Validator[13]. Once the data are organized, they can either be uploaded to an S3 bucket and processed with a command similar to that in Section A.1.2 (updating the bucket name and path to data on the bucket), or kept locally with the `bucket` and `remote_path` values omitted, if one wishes to run the pipeline locally.

## B.3  Changing the Parameters

All of the code for this project is open-source and resides in a Github repository[14]. To test the pipeline with different sets of parameters, it can be cloned and the source code can be modified directly. The repository can be cloned to the `HOME` directory with the following.

```
$ git clone https://github.com/neurodata/ndmg ~/ndmg
```

Once adjustments have been made and the new pipeline is ready to be tested, the package can be re-installed by executing the `setup.py` file contained within the repository.

```
$ cd ~/ndmg
$ python setup.py install
```

## B.4  Changing the Functions

Much like changing parameters, once the repository is cloned it is possible to swap out algorithms or implementations for various parts of the pipeline. Examples of tools which could be replaced include registration or tractography. Again, once this is completed, the pipeline must be re-installed prior to execution.

# Notes

[1] https://www.nitrc.org/forum/forum.php?forum_id=3664
[2] https://cloud.docker.com
[3] http://fcon_1000.projects.nitrc.org/indi/CoRR/html/nki_1.html
[4] Enumerated here: http://bids-apps.neuroimaging.io/apps/
[5] http://brainx.io
[6] https://neurographics.net/2016/07/28/oden-2016/
[7] https://goo.gl/DDMcMG
[8] https://www.cygwin.com/
[9] https://aws.amazon.com/batch/
[10] https://github.com/neurodata/sic/tree/master/code/ec2/batch/json_files
[11] https://github.com/neurodata/sic/blob/master/code/ec2/batch/json_files/job.json
[12] https://www.docker.com/products/overview
[13] http://incf.github.io/bids-validator/
[14] https://github.com/neurodata/ndmg

NeuroData